# Incoherent light induced self-organization of molecules

S. Ahmadi Kandjani

*POMA UMR 6136; Université d'Angers, 2 Boulevard Lavoisier, 49045 Angers, France, and*

*RIAPA, University of Tabriz, 51664 Tabriz, Iran*

R. Barille, S. Dabos-Seignon, J.-M. Nunzi[*]

*POMA UMR 6136, Université d'Angers,*

*2 Boulevard Lavoisier, 49045 Angers, France*

E. Ortyl, S. Kucharski

*Institute of organic and polymer technology,*

*Wroclaw Technical University, 50-370, Wroclaw, Poland*

**ABSTRACT:**

Although coherent light is usually required for the self-organization of regular spatial patterns from optical beams, we show that peculiar light matter interaction can break this evidence. In the traditional method to record laser-induced periodic surface structures, a light intensity distribution is produced at the surface of a polymer film by an interference between two coherent optical beams. We report on the self-organization followed by propagation of a surface relief pattern. It is induced in a polymer film by using a low-power and small-size coherent beam assisted by a high-power and large-size incoherent and unpolarized beam. We demonstrate that we can obtain large size and well organized patterns starting from a dissipative interaction. Our experiments open new directions to improve optical processing systems.

---

[*] Corresponding author : Jean-Michel Nunzi, Université d'Angers, 2 Boulevard Lavoisier, 49045 Angers, France, tel : +332 4173 5364, fax : +332 4173 5216, e-mail : jean-michel.nunzi@univ-angers.fr

# Incoherent light induced self-organization of molecules

S. Ahmadi Kandjani, R. Barille, S. Dabos-Seignon, J.-M. Nunzi, E. Ortyl, S. Kucharski

We contradict the general opinion about surface relief gratings (SRG) that patterning of photoactive systems requires coherent light [1]. It relies on the common sense that interferences and therefore coherence of the interacting waves is mandatory for long-range ordering [2]. Some molecules move under light excitation [3]. Motion is not an organized process, but if the molecules exchange position information, they can organize into well-defined structures. Spontaneous pattern formation arises from the interplay between self-action (e.g. photoisomerization [4]) and long-range interaction (e.g. diffraction [5]). The phenomenon is initiated by noisy random fluctuations. The perturbations bearing a particular periodicity are enhanced, forming regular patterns [6], according to the "winner takes all" theory [7]. Self-organized photoinduced structures find applications in optoelectronic devices [8]. It was previously recognized that self-organization can occur directly from the short range interaction between molecules [9]. Self-organized optical patterns are also induced from incoherent light beams in photorefractive crystals [10, 11]. More generally, feedback effects induce self organization into well ordered patterns in all kind of dynamic systems [12].

In the standard model of SRG formation in azo-dye based photoactive polymers, the photoactive molecules are excited by the illumination pattern. The highly reactive molecules move in a non-uniform way, inducing a mass transport from the bright to the neighboring dark regions [1]. The maximum heights of the light induced SRG correspond to light intensity minima. In this report, we show that a well defined SRG is induced in an azo-polymer film by the combination of a low power coherent laser beam with another high power incoherent and unpolarized beam. It appears that the low power beam carries information about pitch and orientation of the diffraction grating which is transferred to the high power beam. This one creates the structure. We find in this way a simple system allowing to figure out the minimal requirements to organize disordered materials into well organized structures. We verify experimentally that random motion plus information exchange lead to self-organization. Information is exchanged locally via the relief of the polymer surface.

Our samples are polymer films made from a photoactive azobenzene derivative containing heterocyclic sulfonamide moieties [13]. 1 µm-thickness polymer films on glass substrates were spin-coated from a tetrahydrofuran solution. The $\lambda$ = 476.5 nm laser line of a continuous argon ion laser excites the azo-polymer close to its absorption maximum. Experimentally, the laser beam is divided into two arms of a Mach-Zehnder interferometer. First arm carries the seed. Second one carries the totally incoherent and unpolarized pump beam. Incoherence is provided by focusing the laser beam through a diffuser. A hot-air gun used as a turbulent layer generator is inserted along its path in order to insure total incoherence and depolarization [14]. We checked that interferences from a Young's slit setup had zero visbility. Seed and pump beams overlap spatially at the exit of the interferometer, with the thinner seed aligned in the centre of the pump. Both beams illuminate the polymer film from the polymer side. Seed has 0.7 mW-power, 2.1 mm-diameter and linear polarization. Pump has 87 mW-power and 5.8 mm-diameter. Two series of experiments have been performed.

In the first series, the seed is firstly sent alone onto the sample without the incoherent beam. Laser intensity is chosen in order to excite the polymer below the threshold for inducing a SRG on the film. We have evidence that a SRG is induced by measuring the first order diffraction. By a careful control of the diffraction intensity, we have checked that no effect was produced on the polymer surface after more than one hour illumination with 0.7 mW power. Secondly, the unpolarized incoherent pump beam is sent alone onto the film. No SRG is induced and no diffraction detected, even with a longer exposure time. Thirdly, when the two beams are sent together and overlap onto the polymer film, a SRG is induced after ½ hour exposure. Both the height and the pitch of the grating were retrieved using a contact-mode atomic force microscope (AFM). A well defined SRG is induced in the overlapping region of the coherent seed (figure 1a). The structures observed in the non-overlapping region (figure 1b) reflect amplified fluctuations due to the random molecular movements photo-induced by the incoherent beam. The peak-to-peak height of the grating in figure 1a is $100 \pm 5$ nm. Its mean pitch is $\Lambda = 890 \pm 30$ nm. Grating wave-vector is parallel to the linear polarization of the seed, which is in the *p*-plane. The pitch is close to the theoretical value given by first order diffraction theory in the backward direction $\theta = 32.6°$: $\Lambda = 2 \lambda / 2 \sin \theta$. Far outside of the coherent beam region, no grating is induced after ½ hour illumination.

In the second series of experiments, the coherent seed is firstly sent alone on the polymer with a large power of 125 mW during one hour. A well-defined SRG is stored. Its spatial frequency profile is shown as a continuous line in figure 2d. Secondly, the coherent beam power is decreased to 0.7 mW. The coherent beam overlaps partially with the totally incoherent beam (scheme in figure 2b). After 2 hours exposure, a well-defined SRG is also stored (figure 2a). Importantly, the amplitude profile of the SRG taken in all places inside the incoherent beam region has the same characteristics in height, pitch and orientation as the one induced by the high power coherent beam alone. This is confirmed by the Fourier transform given in figure 3d. It shows that the two patterns have the same spatial frequency within experimental uncertainties. So, the SRG has expanded from the central coherent beam region to the whole region exposed to the incoherent beam. In this way, we see that a totally incoherent beam can provide the movement which is necessary to induce a well-defined SRG.

The information brought by the coherent seed has been transmitted to peripheral incoherent region by the molecular self-assembling process: ie. the organized molecules communicate non-local information about photo-induced structural organization to the non-organized neighboring ones. We also checked that before reaching saturation, the self-structuring process which occurs under incoherent illumination continues up to saturation after removal of the seed. Moreover, if the incoherent beam is polarized, the polarization of the seed still governs the orientation of the structure. No structure appears out of the incoherently illuminated region.

The origin of the driving force responsible for SRG formation is summarized in the following. As a result of isomerization, the molecules migrate along their main axis, almost parallel to the polarization direction [15]. They diffuse from illuminated to dark regions and get trapped into the dark regions. In particular, when a collimated laser beam impinges onto the sample, light is diffused by the polymer film according to Huygens principle, as a spherical wave in all directions around micro-roughnesses. This creates light intensity fluctuations: there is fewer light below the roughness peaks and more light in the near vicinity around the peaks. Molecule migration amplifies the roughness into surface profile fluctuations. Some diffracted light will interfere constructively in the diffraction direction [5]. This will order the surface profile fluctuations into a periodic pattern. The process will self-saturate when a balance between incident and diffracted light is reached, following coupled wave theory [16]. The process which starts from random fluctuations will converge into a well-organized SRG initialized by the information contained into the coherent seed. We illuminated the molecules with totally incoherent light and added a coherent seed. The molecules which respond to the coherent seed create a diffraction pattern and then transmit the pattern step by step to all their neighbours. Information is exchanged locally via the relief: pits sending more light to the neighboring and dips sending less light, thus replicating the pattern. As a consequence, temporal coherence between pump and seed should not be necessary, but little spatial coherence of the incoherent beam is necessary within a few micrometer length scale.

In conclusion we have shown that a well-defined pattern can be printed in an azo-polymer film illuminated by a large power incoherent beam. Information about the structure is brought by a weak confined laser beam. Our experiments suggest applications in optoelectronics in which a low-power compact laser source can be used to encode information which will be transferred to a polymer film assisted by a strong incoherent light source. Our experiment shows that complex behavior can be experimented using simple systems: weak coherent light can serve as a seed to create information into a polymer film in such a way that molecules powered by incoherent light will build and transmit well defined patterns. Interestingly, we see that a complex structure is built from moving objects although the structure is ignored at the individual level. Further study will be devoted to the investigation of the nonlinear behavior of the coherent-incoherent beam interaction leading to pattern formation: non-local, threshold, coherence and temporal effects.

**REFERENCES**

[1] C. Cojocariu, P. Rochon, "Tribute to Almeria Nathanson: Light-induced motions in azobenzene-containing polymer", *Pure Appl. Chem*. **76**, 1479 (2004)

[2] M.C. Cross, P.C. Hohenberg, "Pattern formation outside of equilibrium", *Review of modern physics* **65**, 851 (1993)

[3] Y Tabe, T Yamamoto, H Yokoyama, "Photo-induced travelling waves in condensed Langmuir monolayers", *New J. Phys.* **5,** 65 (2003)

[4] C. Jones, S. Day, "Shedding light on alignment", *Nature* **351**, 15 (1991)

[5] A.E. Siegman, P.M. Fauchet, "Stimulated Wood's anomalies on laser-illuminated surfaces", *IEEE J. Quant. Electr.* **22**, 1384 (1986)

[6] C. Hubert, C. Fiorini-Debuisschert, I. Maurin, J.M. Nunzi, P. Raimond, "Spontaneous patterning of hexagonal structures in an azo-polymer using light-controlled mass transport", *Adv. Mater*. **14**, 729 (2002)

[7] W.J. Firth, "Pattern formation in passive nonlinear optical systems" in *Self-Organization in optical systems and applications to information technology*, M. A. Vorontsov, W. B. Miller, eds. (Springer, Berlin), 69 (1995)

[8] A. Natansohn, P. Rochon, "Photoinduced motions in azobenzene-based amorphous polymers: possible photonic devices", *Adv. Mater*., **11**, 1387 (1999)

[9] W. Lu, D. Salac, "Patterning Multilayers of Molecules via Self-Organization", *Phys. Rev. Lett*. **94**, 146103 (2005).

[10] D. Kip, M. Soljacic, M. Segev, E. Eugenieva, D.N. Christodoulides, "Modulation Instability and Pattern Formation in Spatially Incoherent Light Beams", *Science*, **290**, 495 (2000).

[11] T. Schwartz, T. Carmon, H. Buljan, M. Segev, "Spontaneous Pattern Formation with Incoherent White Light", *Phys. Rev. Lett*., **93**, 223901 (2004).

[12] A. S. Mikhailov, V. Calenbuhr, "From Cells to Societies: Models of Complex Coherent Action" (Springer, Berlin 2002)

[13] E. Ortyl, R. Janik, S. Kurcharski, "Methylacrylate polymers with photochromic side chains containing heterocyclic sulfonamide substituted azobenzene", *Eur. Polymer J.*, **38**, 1871 (2002)

[14] A. Consortini, Y.Y. Sun, C. Innocenti, Z.P. Li, "Measuring inner scale of atmospheric turbulence by angle of arrival and scintillation", *Opt. Comm*. **216**, 19 (2003)

[15] P. Lefin, C. Fiorini, J.M. Nunzi, « Anisotropy of the photoinduced translation diffusion of azobenzene-dyes in polymer matrices », *Pure Appl. Opt.* **7**, 71 (1998)

[16] H. Kogelnik, "Coupled wave theory for thick hologram gratings", *Bell Sys. Technol. J.* **48**, 2909 (1969)

**Figure captions**

**Figure 1**: AFM images of self-patterned SRG structures obtained after 30 mn exposure with a low power coherent beam and a large power incoherent beam.

a) We scan the SRG in the central part of the coherent beam region. Isomerization induces a molecular migration almost parallel to the polarization direction. In particular, when a collimated laser beam impinges onto the polymer surface, light is diffused inside the polymer film in all directions around any micro-roughness. It will diffract a light amplitude into the film plane. The roughness amplitude is increased. This creates a surface modulation which organizes coherently to diffract the incident laser light out of the polymer film. The modulation amplitude increases with time from its initial value, up to saturation of the diffraction efficiency. Saturation corresponds to the balance between incident and diffracted intensities: more light coupled by diffraction into the polymer film resulting into more light diffracted out of the polymer film. Backward diffracted intensity experimentally reaches about 1%.

b) Within the outer incoherent laser beam region, light induces a random motion of the molecules. No well defined pattern can be produced. Within short exposures, roughness amplitude is increased, without any coherent coupling.

**Figure 2**: Self-patterned SRGs obtained with a totally incoherent beam after 2 hours of exposure.

a) AFM image of the self-patterned SRG induced by the incoherent beam, outside of the coherent beam region. This pattern shows that the information about the structure was transmitted from the coherent to the incoherent regions.

b) Sketch of the expansion mechanism showing propagation of the structure from the coherent to the incoherent region, up to the whole illuminated area.

c) The amplitude profile of the grating taken along the yellow line in the SRG pattern shows the regularity of the profile. Grating amplitude is 100 nm, pitch is 890 nm.

d) A plot of the spatial frequency pattern allows comparison of the SRGs formed using a strong coherent beam alone and in the region outside of the coherent beam in the dual beam experiment of part b.

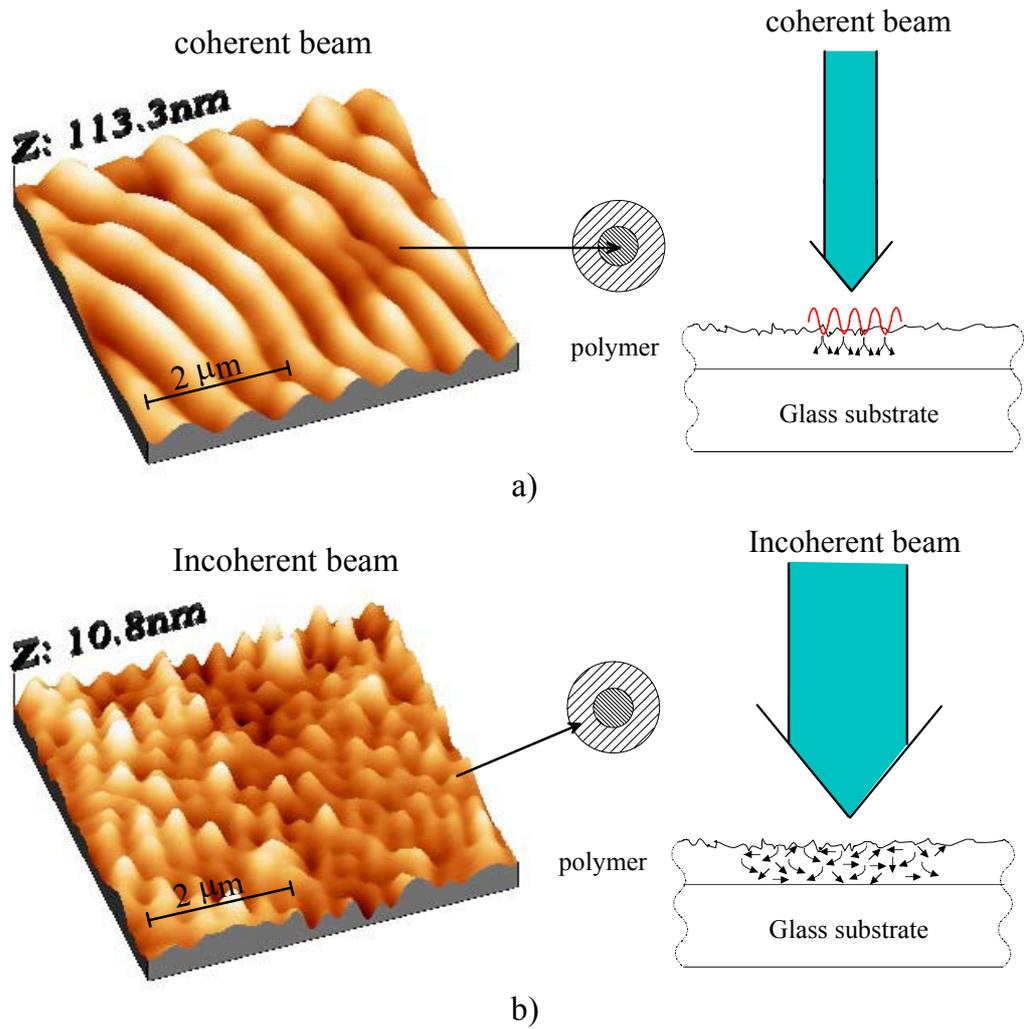

**Figure 1**

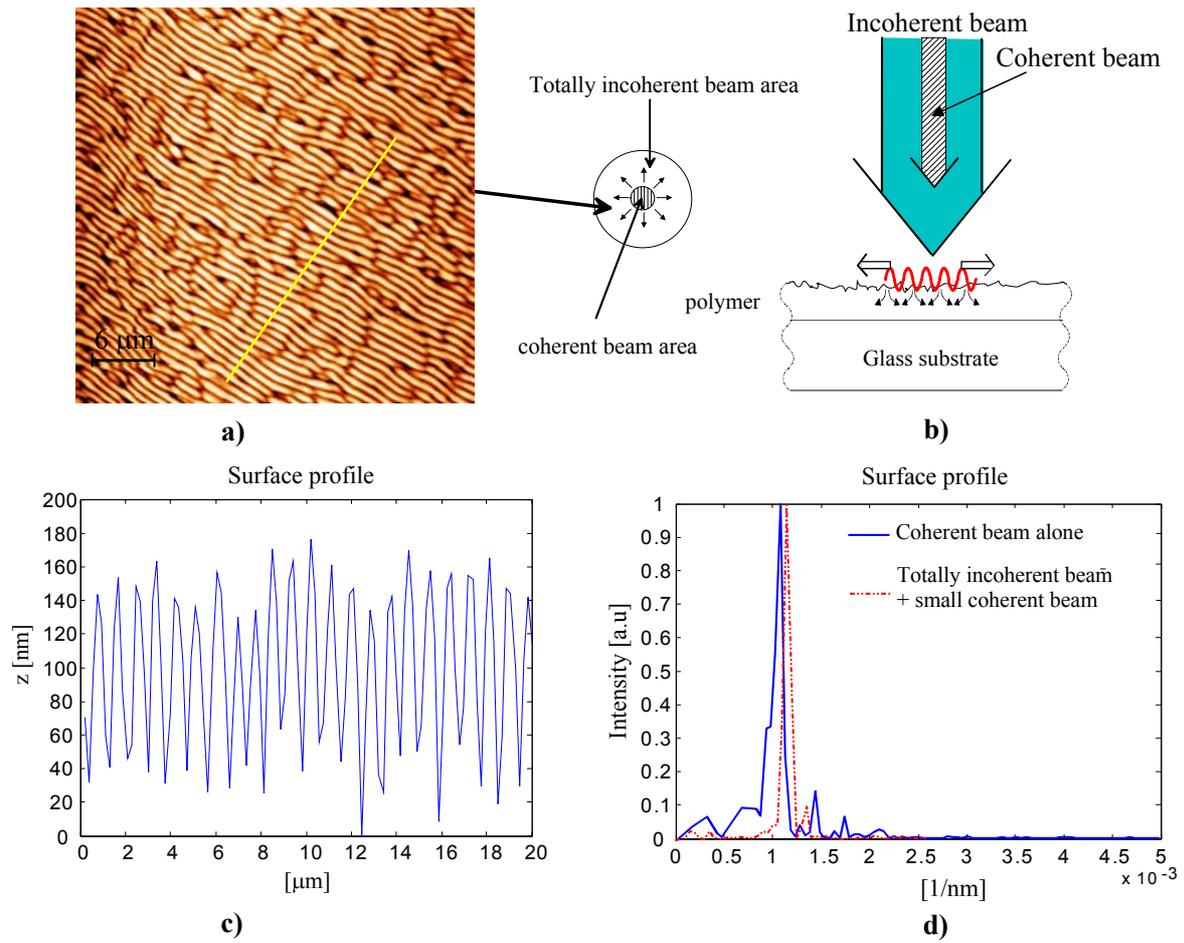

**Figure 2**